---------------------------------------------------------------
\documentstyle[twocolumn,prl,aps]{revtex}
\begin{document}
\draft
\title{ {\large  to be published in Phys. Rev. Lett. 1995 (e-print
archive
chao-dyn@xyz.lanl.gov,9501006 )}\\[10mm]
Formation of Dynamic Domains in Strongly Driven Ferromagnets}
\author{T. Plefka}
\address{ Theoretische Festk\"orperphysik, Technische Hochschule
Darmstadt,
D 64289  Darmstadt, Germany}
\date{December 1994}
\maketitle
\begin{abstract}
Based on the dissipative Landau-Lifshitz equation,
the spatiotemporal structure formation problem
is investigated in the
region far above the transverse ferromagnetic resonance instability.
Apart from the external fields, the model
contains an isotropic exchange field, a shape demagnetisation field
and an anisotropy field. Numerical simulations
exhibit in the rotating frame a stationary domain structure
with  a precessing motion  in the walls regimes. Employing
analytical methods,  characteristic elements of this
structure are explained. This driven
dissipative system shows similarities to equilibrium systems
of coexisting phases and   organises itself in such a
way that the local dynamics tends to become hamiltonian.
\end{abstract}
Ferromagnetic systems strongly driven by external transverse
magnetic fields have been under investigation for more than four
decades (for the early work see \cite{da63}). Based on a
nonlinear spin-wave expansion around the homogeneous ground
state, Suhl \cite{su57} showed that certain modes become
unstable for increasing amplitude of the pump field. Detailed
measurements exhibit beyond threshold many of
the phenomena predicted by the general theory of non-linear
dynamics. Based on \cite{su57}, truncated spin-wave mode models
have been published, which  successfully
describe the experimental findings (for a recent,
comprehensive presentation see \cite{wigen94}).

Apart from single exceptions like
the simulations of Elmer \cite{el88}, who found
dynamic domains for a special model, the existing
investigations are limited to the weak nonlinear regime.
The regime far above the instabilities with pump
amplitudes of the order of the demagnetisation field
remains an open question (compare \cite{cr93} p. 1075).

This  question is of high interest in the
general theory of spatiotemporal pattern formation
in dissipative systems \cite{cr93}. At the beginnings
of this  modern and interdisciplinary field of research
Anderson \cite{an81} proposed that a driven ferromagnet
is the characteristic example for pattern formation and for the
general problem of how the concepts of equilibrium phase
transitions can be extended to driven dissipative systems.
Recently, Cross and Hohenberg \cite{cr93} raised
doubts against this proposal.

It is mainly the dipole interaction which makes
pattern formation in ferromagnetic materials a difficult
problem. The difficulties already arise in the undriven
case \cite{doering66}. In a first step, the theory of static
ferromagnetic structures \cite{dom} usually employs
an approximation. Above all, stray fields are omitted
and only the part of the dipole interaction
which describes the shape demagnetisation fields is considered. Even
in this approximation many interesting results have been
obtained \cite{dom} both for the formation of
static structures and for the wall dynamics.

Using this approximation for the dipole field, this work
adresses  the
pattern formation problem in driven magnetic systems.
 Focusing on the region far above threshold,
 it is the aim of the present investigations to
work out general features and mechanisms by employing
both analytical and numerical methods. It is beyond
the scope of this paper to improve the existing near
threshold treatments, as this would certainly require
the complete dipole field
used in the spin-wave approaches.

At a mesoscopic scale the dynamics is governed by the
Landau-Lifshitz equation as was recently demonstrated by microscopic
investigations \cite{pl90,ga91}. In the frame rotating with
the driving frequency
$ \omega $ around the $ {\bf e}_z $ direction,
this equation of motion takes the form
\begin{equation} \label{1}
{\bf \dot{m}}\,=-{\bf m\times}
\biglb( {\bf H}^{{\rm eff}}\,-  \omega {\bf e}_z  \biglb)\,-
\Gamma \,{\bf m\times \bigl( m \times H }^{{\rm eff}} {\bf
\bigl)}\,.
\end{equation}
$ {\bf m}({\bf r},t) $ and ${\bf H}^{{\rm eff}} $
are the local magnetisation and the effective
field in the rotating frame, being related to the quantities $
{\bf m}_{{\rm lab}} $ and ${\bf H}^{{\rm eff}}_{{\rm lab}} $
in the laboratory frame by $ {\bf m}_{{\rm lab}}\,=\, \exp
(\,  \omega  t {\bf e}_z \times \,)\,\,{\bf m} $ and by $ {\bf
H}^{{\rm eff}}_{{\rm lab}}\,=\, \exp (\, \omega  t {\bf e}_z \times
\,) \,{\bf H}^{{\rm eff}} $, respectively. $ \Gamma $
represents the Landau-Lifshitz damping rate. In reduced units
the gyromagnetic ratio and the magnitude of the magnetisation
$m\,=\, |{\bf m}|$ are equal to 1.

Specifying the  model under investigation,
it is assumed for the effective field that
\begin{equation} \label{2}
{\bf H}^{{\rm eff}}\,= h_\parallel {\bf e}_z \,
 + \, h_\perp{\bf e}_x\,- \,
\bbox{\overline{m}} \,+\, J  \Delta {\bf m}
+A m_z {\bf e}_z
\end{equation}
where $ h_\parallel $  and  $ h_\perp $ are the
amplitudes of the external static and the external
circular driving rf field,  respectively. The term $
\bbox{\overline{m}}\,=
\,V^{-1}\int{\bf m}\,{\rm d}V $ represents, again in reduced
units, the demagnetisation field of a sphere of volume $ V $.
The contribution  $ J  \Delta {\bf m}$  results from the isotropic
ferromagnetic exchange interaction. An uniaxial  anisotropy is described
by $ A m_z {\bf e}_z  $.

In the first part of this letter the results of
numerical simulations in one spatial dimension with
an arbitary $ \xi $ direction
 $ {\bf m}({\bf r},t)\rightarrow {\bf m}(\xi,t) $
are reported using periodic boundary conditions
$ {\bf m}(\xi,t)={\bf m}(\xi+L,t) $. These investigations
were performed on  a vectorized super computer.
The program uses an Euler
integration scheme in time and
is based on semi-spectral methods.

These simulations exhibit a
temporal evolution of $ {\bf m}(\xi,t)$
which is characteristic  for structure formation
in dissipative systems. After a period of transient
behaviour,  which strongly
depends on the special initial
state and which may be very complex, a tendency toward the
formation of domains  is found.
In the early stage these domains
interact which each other, merging into
domains of larger size. With increasing time and
increasing domain sizes,  this process slows
down and finally the structure
becomes stationary on large time scales.

Fig.1 shows such a final  domain state obtained
by a long-time study starting from a randomly disturbed,
homogeneous initial state. The magnetisation
is nearly everywhere constant,
taking the  values
$ {\bf m_+} $  or $ {\bf m_-} $, respectively.
Contrary to this behaviour within the domains,
a strong time dependency is found in  the
narrow  wall regimes which
separate the domains. As  Fig.2
and Fig.3 demonstrate, the magnetisation
$ {\bf m} $ at a fixed position
is anharmonically oscillating with
a characteristic internal period  of
rotation $ T_{\rm per} $, which
is found to be independent of the
specific position. This spatiotemporal
wall structure  is dynamically very stable.
In the simulations no changes of any
significance could be found for
times corresponding to $ 10^4 $ periods
of $ T_{\rm per} $.

It should be pointed out that this wall structure
as well as the macroscopic  quantities
$ {\bf m_+} $  , $ {\bf m_-} $,
$ \bbox{\overline{m}} $  and $ T_{\rm per} $
are independent of the specific initial state
and  thus are characteristic elements for
the structure formation.

To analyse  these characteristic features
found in the simulations
a perturbative approach has
been worked out using the
multiple-time scaling method
and treating
the contributions $ J  \Delta {\bf m} $
and $ A m_z {\bf e}_z $ as perturbations \cite{com1}.
The theory has been worked out up to the
first order. The first order treatment is lengthy
and rather technical \cite{pl95}, thus I restrict myself
here to the zeroth order.

In this order eqs.(\ref 1) and (\ref 2) reduce
to
\begin{equation} \label{3}
{\bf \dot{m}}({\bf r,t})\,=-{\bf m\times} {\bf H}_1\,-
\Gamma \,{\bf m\times \bigl( m \times H }_2 {\bf
\bigl)}\,.
\end{equation}
with
\begin{equation} \label{4}
{\bf H}_2={\bf H}_1+ \omega {\bf e}_z
=h_\parallel {\bf e}_z \,
 + \, h_\perp{\bf e}_x\,- \,
\bbox{\overline{m}}(t) .
\end{equation}
Due to the $ \bbox{\overline{m}} $
term, the problem
described by eqs.(\ref 3) and (\ref 4) is
of the mean field type and the usual technique
can be applied. For the present case this
would imply   solving
eqs.(\ref 3)  for given
$ \bbox{\overline{m}}(t) $ as a first step.
Integration of the obtained solution
$ {\bf m}({\bf r},t) $
over the sample volume V would then lead to a
self-consistency condition for  $ \bbox{\overline{m}}(t) $,
from which $ \bbox{\overline{m}}(t) $
and the complete solution can in principle be
obtained. This procedure can in general not
be performed, as already
the analytic solution of the first step is not
known \cite{ma86}. The fixed points of eqs.(\ref 3),
however, and their stability can explicitly be
determined by applying  the mean field procedure.

The results of this, basically straightforward,
fixed point analysis show
that marginally stable solutions
exhibiting a domain
state structure are possible. The local magnetisation
$ {\bf m}({\bf r}) $  takes only two values
$ {\bf m}_+$ and $ {\bf m}_-$ realised in the
generally disconnected partial volumes
$V_+=n_+ V $ and $ V_-=n_-V$, respectively.

In such  domain states the fields
$ {\bf H}_1 $ and $ {\bf H}_2 $
become orthogonal
\begin{equation} \label{5}
 {\bf H}_1{\bf H}_2= 0
\end{equation}
and it is advantageous to write the
results in terms of the internal coordinates introduced
by  $ {\bf e}_1={\bf H}_1/H_1\,, {\bf e}_2={\bf
H}_2/H_2 $ and by $ {\bf e}_3={\bf e}_1{\bf \times e}_2 $.
The transformation from the internal to the primary
frame is found to be
\begin{equation} \label{10}
 \left(\matrix{{\bf e}_x \cr {\bf e}_y \cr {\bf e}_z} \right)
= \frac{1}{u w} \left(
          \matrix{
                -d h_\perp &         -d u    & \Gamma w \cr
     \Gamma h_\perp & \Gamma u & d w \cr
         -u^2                  &   u h_\perp & 0}
    \right)
 \left(\matrix{
{\bf e}_1 \cr{\bf e}_2 \cr{\bf e}_3} \right)
\end{equation}
where
\begin{eqnarray}
\label{6}
d & = & h_\parallel -\omega \; ; \quad
u  =  ( d^2+\Gamma^2)^{1/2} \\*
v &  = &
 ( u^2-\Gamma^2 h_\perp^2)^{1/2}\; ; \quad
w   =  ( u^2+h_\perp^2)^{1/2}
\end{eqnarray}
were introduced. The magnitudes of the fields $ {\bf H}_1 $
and  $ {\bf H}_2 $ are calculated to take the values
\begin{equation}
\label{77}  H_1 =\omega u w^{-1} \quad ;\quad
            H_2 =\omega h_\perp w^{-1}
\end{equation}
and the domain magnetisations are found to
be given by
\begin{equation}
\label{7} {\bf m}_\pm =\;u^{-1} ( \mp v\, {\bf e}_1 \;
+\; \Gamma h_\perp \,{\bf e}_3 ).
\end{equation}
For the bulk magnetisation in a domain state,
the self-consistency condition leads  to
\begin{equation} \label{8}
{\bf \overline{m}}_{{\rm dom}}=
n_+ {\bf m}_+ \;+\;n_- {\bf m}_-
\end{equation}
and the $ n_\pm $ are calculated to be
\begin{equation} \label{9}
n_\pm=\; \frac{1}{2} \; \pm \frac{d w^2 + \omega u^2}{2 w v}.
\end{equation}
The domain states exist only in a
subspace of the parameter space spanned by
$ h_\parallel, h_\perp,\omega $ and by $ \Gamma $.
This subspace is implicitly
determined by the condition  $ 0< n_+ <1 $.
For parameter values outside this subspace
eq.(\ref3) has a different stable fixed point solution
characterised by a homogeneous magnetisation
$ {\bf m}({\bf r})={\bf m}_{\rm hom} $. These
homogeneous states, being of
minor interest for the present work,
are unstable in the regime where
the domain states exist.
Apart from the domain and
the homogeneous states, no other stable
fixed points of eq.(\ref3)  exist.

Next, the findings of the linear stability analysis
about the domain state fixed points are presented.
Employing the usual $ \exp(\lambda t) $ ansatz for
the deviations, one finds for
all modes with the exception of four
\begin{equation}
\lambda = \pm i \;\Omega \quad {\rm with}\quad
\Omega= \omega v w^{-1} \label{21}
\end{equation}
which implies that they are undamped and
oscillating with $\Omega$.
The four remaining modes are of collective
character as they describe
disturbances
of the form $  V^{-1}\int_{V_\pm}
({\bf m} ({\bf r},t)\,-\,{\bf m}_\pm)\;{\rm d}V $.
The characteristic equation for these
collective modes is found to be given by
\begin{eqnarray}
0 & = & |\Lambda_+ \Lambda_-|^2
-  2  n_+  n_- v^4 u^{-4}
\bigl[ \Re( \Lambda_+
\Lambda_-^\ast)-n_+  n_- \bigr]
\nonumber \\
& - & n_+^2 n_-^2
 -  2 n_+ n_- \Gamma^4 h_\perp^4 u^{-4}\bigl
[\Re( \Lambda_+ \Lambda_-)-n_+ n_- \bigr],
  \label{22}
\end{eqnarray}
where $ \Lambda_\pm =(\lambda\mp
i \Omega)/(\Gamma +i) + n_\pm $ was introduced. It has been shown
analytically for the weak damping case ($ \Gamma \ll 1$)
and numerically for the general case,
that the collective modes are damped everywhere.
Note
that this implies relaxation
of $ \bbox{\overline{m}}(t) $ to
$ \bbox{\overline{m}}_{\rm dom} $.

Under the constraint  $ \bbox{\overline{m}}(t)
= \bbox{\overline{m}}_{\rm dom} $, the analysis
of eqs.(\ref{3}) and (\ref{4}) can be extended
to the nonlinear regime.
Employing eq.(\ref{77}), this leads to
\begin{equation} \label{23}
{\bf \dot{m}}\,=-\omega w^{-1}\Bigl(u\; {\bf m\times} {\bf e}_1\,+
\Gamma h_\perp \,{\bf m\times \bigl( m \times e }_2 {\bf
\bigl)} \Bigr)\,.
\end{equation}
For this equation of motion
the quantity
\begin{equation} \label{24}
m_1({\bf r},t) (\Gamma h_\perp
m_3({\bf r},t)- u)^{-1}= v C({\bf r})
\end{equation}
is a constant of motion and  eq.(\ref{23})
is integrable, as in addition
$ m_1^2+m_2^2+m_3^2=1 $ holds.
Eq.(\ref{24}) represents a plane  and thus the resulting
motion of ${\bf m}({\bf r},t) $ is a precession
on a cone at an arbitrary
but fixed position $ {\bf r} $.
This precession is found
to be anharmonic with the period
of rotation of $  T_{\rm per}= 2 \pi
\Omega^{-1} $ independent of
$ {\bf r} $.

The values of $ C({\bf r}) $ are restricted
by the condition $ |C({\bf r})|\le 1 $, as otherwise
the plane (\ref{24}) does not intersect with the sphere.
For the limiting values $ C({\bf r})=\pm 1 $ the plane becomes
tangential and
consequently $ {\bf m} $ becomes time independent
taking  the fixed point values $ {\bf m}_\pm $ of eq.(\ref{7}).
Thus a wall
is characterised by a change from
$ C=-1 $ to $ C=1 $. Considering now the
$ V \rightarrow \infty $
limit, the constraint $ \bbox{\overline{m}}(t)
= \bbox{\overline{m}}_{\rm dom} $ can
be satisfied, as long as the entire wall volume goes to zero
faster than $ V $.

Summing up the analysis of eqs.(\ref{3}) and (\ref{4})
has shown that  stable  solutions of
domain type  do exist. In the domain regions
the local magnetisation is stationary but precesses
with $ T_{\rm per} $ in the wall regime. It is obvious that these
results
are  qualitatively in accord with the  numerical findings.
Using the parameter value of the simulation,
quantitative agreement is found for
$ \bbox{\overline{m}}_{\rm dom} $, for
$ \bbox{m}_\pm $, for $ n_\pm $ and for
$ T_{\rm per} $. The deviations are less than one percent.
Recall also that the numerical
results of Figs.1-3 were already  presented  in terms of
the internal coordinates (\ref{10}). Furthermore, the numerical
results show that the motion in the wall regime is
planar and well described by eq.(\ref{24}). Finally,
the first order treatment  \cite{pl95} of
eqs.(\ref{1}) and (\ref{2}) is able to explain  with high
accuracy  the
complete spatiotemporal wall structure
of Figs.2 and 3.

Some features of the present investigation
which may be of general interest should be pointed
out.

The behaviour of this dissipative system driven away from
equilibrium  in the stationary domain state is
analogous to a thermodynamic
system {\it at} a first order equilibrium
transition exhibiting coexisting phases. The translational
symmetry is broken for the domain structure.
Macroscopic variables, as in the present case
$ \bbox{\overline{m}}_{\rm dom} $, $ \bbox{\overline{m}}_\pm $
and $ n_\pm $, are independent of the initial conditions
and only depend on the external parameters like $ h_\parallel$,
$ h_\perp $ and $ \omega $. In this context eq.(\ref{5})
has to be interpreted as the criterium for existence of the
domain states.
Note that this criterium reduces for the
static case ($\omega=0$) to $h_\parallel {\bf e}_z \,
 + \, h_\perp{\bf e}_x\,- \,
\bbox{\overline{m}}_{\rm dom}=0 $,
which  means that the internal field  vanishes.
As this is the usual criterium for a ferromagnet to be at
the equilibrium phase transition, some aspects of the
present work  can be interpreted as an extension of the
concepts of phase transitions to driven systems.

Another  interesting aspect of the present work is the
local generally nonlinear dynamics within the domain regimes. As long
as the sample is sufficiently large, this local dynamics
is integrable or is hamiltonian according to the investigations
of eq.(\ref{23}). Although the system becomes weakly
dissipative in the first order treatment \cite{pl95},
it is worthwhile to point out
this tendency of the system to  organise  itself on the
global macroscopic scale in such a way
that the local mesoscopic dynamics becomes at least nearly hamiltonian.

In this context it is interesting to go back to the microscopic scale
\cite{pl90}, where the system is a quantum mechanical many
body spin system in contact with a
bath of sufficiently low temperature.
It is basically the competition between the short ranged
exchange interaction and the long ranged repulsive dipole
interaction, which leads to the formation of structure and
to the nearly hamiltonian dynamics on the mesoscopic scale.

Ferromagnets are  representative of a whole
class of many  body systems characterised by
short ranged attractive interactions  in competition
with long ranged  repulsive interactions. Thus, the
question arises whether the tendency toward a local hamiltonian dynamics
is generic for all these  rather common systems.
Should this be the case, it would perhaps become clear
why structure and why hamiltonian dynamics are so widespread in nature.

Coming back to the ferromagnet, it is pointed out that neither the
location nor the geometric shape of the walls is determined by the
present approach. By analogy with the static case \cite{dom}, it is
expected that the dipolar stray field and the usual
boundary effects will reduce this freedom.  The formation of
regular patterns in driven ferromagnets seems to be possible.

For completeness, the results of this work
are specialised to the weakly
nonlinear regime $ h_\perp\ll1$. For simplicity, the discussion will
be restricted to $ \omega = h_\parallel $ and to $ \Gamma \ll 1$. A
critical value
of the pump field $ h^{\rm crit}= \Gamma (h_\parallel ^2-1)^{1/2}$
results. For $ h_\perp <h^{\rm crit} $ the  homogeneous state is
stable and
for $ h_\perp >h^{\rm crit} $ the domain states arise. As
in this work the dipolar stray field is neglected,
$ h^{\rm crit} $ differs from the threshold values of \cite{su57},
which under usual conditions are smaller than $ h^{\rm crit} $.
Lastly, the findings for the power absorption $P$
will be given. For the homogeneous state
$ P= \Gamma ^{-1}
h_\perp^2 $ and for the domain states
$ P = \Gamma h_\parallel^2 h_\perp^2 (\Gamma^2 + h_\perp^2 )^{-1} $
are obtained. This result implies an  experimentally well established
\cite{da63,wigen94}
saturation of $ P $  for $ h_\perp \gg \Gamma $. Thus, as a side product
of this work it must be concluded that this
saturation is  not solely explained through spin-wave approaches.

I am particularly indebted to F. Matth\"aus for his
program from which the numerical results
have been obtained. I gratefully acknowledge discussions with
H. Benner, H. J. Elmer, W. Just and G. Sauermann. This work
was performed within SFB 185.

\begin{figure}
\caption{Spatial dependence of a stationary domain state in the rotating
frame. Plotted are $ \cos \phi = m_1 ( m_1^2 +m_2^2 ) ^{-1/2}$
and $m_3$, where $ m_i $ are the components of the local magnetisation
in the internal coordinates, defined by eq.(6). The simulation
uses $1024$ mesh points and the  parameter values
$ h_\parallel=\omega=2,\; h_\perp=.5,\; J=.01,\;A=-.005,\;\Gamma=.1 $
and $ L=1024 $.}
\label{fig1}
\end{figure}
\begin{figure}
\caption{Spatiotemporal structure of $\cos \phi $ in the vicinity
of a wall
located at $ \xi_{\rm wall} $ with $ T_{\rm per}=18.6 $.
Definitions and parameter values as in Fig.1.}
\label{fig2}
\end{figure}
\begin{figure}
\caption{Spatiotemporal structure of $m_3 $ in the vicinity of a wall.
Definitions and parameter values as in Figs.1 and 2.}
\label{fig3}
\end{figure}

\begin{references}
\bibitem{com1} Note that  $ J  \Delta {\bf m} $ is only a secondary
contribution of the strong exchange field. The main
contribution  $ J _0 {\bf m} $ enters only in the magnitude of the
local magnetisation (compare \cite{pl90}), which is masked
as reduced units are used.
\end{references}
\end{document}